 \documentclass[final,3p,times,twocolumn]{elsarticle}

\usepackage{amssymb}
\usepackage{amsmath}
\usepackage{color}

\journal{Physica E: Low-dimensional Systems and Nanostructures}

\begin{document}

\begin{frontmatter}



\title{A Comparative Study of Substrate’s Disorder on  Mobility in the Graphene Nanoribbon: Charged Impurity, Surface Optical Phonon, Surface Roughness}


\author{Shoeib Babaee Touski}
\ead{touski@hut.ac.ir}
\address{Department of Electrical Engineering, Hamedan University of Technology, Hamedan 65155, Iran}
\author{Manouchehr hosseini}
\address{Department of Electrical Engineering, Bu-Ali Sina University, Hamedan, Iran.}

\begin{abstract}
The effects of substrate on the electronic properties of Graphene remains unclear. Many theoretical and experimental efforts have been done to clarify this discrepancy. In this work, we studied the electronic transport in armchair Graphene nanoribbons (AGNR) in the presence of substrate's disorder. The three main substrate's disorders -surface roughness, charged impurity and surface optical phonon- are investigated. Non-Equilibrium Green's function along with the tight-binding model is employed to investigate the electronic properties of Graphene Nanoribbons. The effects of these disorders are investigated individually, finally, the effects of them are compared to determine the dominant source of scattering. 

\end{abstract}


\end{frontmatter}


\section{Introduction}
Two dimensional (2D) materials have attracted great attention due to their interesting properties. Graphene, the first member of this family, exhibits unique electronic properties and high carrier mobility \cite{neto09,bolotin08}. These materials should be placed on a substrate for application in the electronic devices and experimental results indicate that substrate highly affects the electronic properties of 2D materials. Freestanding Graphene mobility is about 200,000 $\mathrm{cm^2/Vs}$ \cite{bolotin08,bolotin08tem,du08} whereas supported Graphene mobility declines to $\mathrm{1000-20000 cm^2/Vs}$ \cite{zhang05,tan07}. These discrepancies are usually explained by interactions of Graphene with underlying substrates.

The main sources of the substrate scattering are substrate impurities \cite{ando06,hwang07,chen08charged}, surface phonon of polar substrate \cite{fratini08,chen08intrinsic,meric08,freitag09} and surface roughness \cite{Morozov08}. The effects of charged impurity (CI) is suggested as the main source of scattering \cite{ando06} but it has been shown that charged impurity scattering in low-dimensional semiconductor nanostructures can be damped by coating them with a high-$K$ dielectrics \cite{jena07}. Experimental results have shown that mobility of Graphene can be increased with putting a high-$K$ dielectric on it \cite{jang08,newaz12,ong13}.
At the other hand, deposition of dielectric films on Graphene has decreased electron mobility due to transferring surface roughness to Graphene \cite{lemme08}. It has been observed that with deposition of ultra-thin high-$K$ dielectric materials on Graphene with much less roughness can improve the carrier mobility \cite{fallahazad10,fallahazad12}. Hollander et al. [\cite{hollander11}] reported a mobility enhancement in Graphene with applying a thinner top gate dielectric. 
The third effect, surface optical phonon (SOP) is transferring to Graphene that can highly scatter carriers \cite{konar10}. SOP scattering from polar dielectric layer can be a dominant factor in limiting carrier mobility of Graphene at room temperature insufficiently clean samples \cite{chen08intrinsic,fratini08}. 

Recently, hexagonal boron-nitride (BN), a 2D crystal with a lattice constant very close to Graphene, has been proved to be a good match substrate owing to its smooth morphology, the limited density of dangling bonds, and a little charge trapping. These characteristics of BN lead to a dramatic improvements in Graphene mobility ($140000 \mathrm{cm^2/Vs}$) \cite{dean10,xue11,mayorov11}.

Surface roughness is proposed as another source of scattering from the substrate that has not been comprehensively investigated. Reducing the surface roughness is critical for Graphene-based devices because local curvatures can lead to electronic effects such as creating charge puddles\cite{guinea08}. The surface roughness of Graphene on the different substrates is experimentally studied \cite{ishigami07,lui09,geringer09,dean10}. In our previous work, we studied the effects of surface roughness on the electronic properties of Graphene nanoribbon \cite{touski13}. It is shown that the mean free path can decrease with the power of four of root mean square (RMS) of surface roughness.

In this paper, we study the effects of charged impurities, surface optical phonon and surface roughness on the mobility of Graphene nanoribbon and compare the effects of them.

\section{Approach}
Experimental results have shown that the underlying substrate can highly affect the electrical properties of Graphene nanoribbon and drastically decline mobility. Between different methods to explain this phenomenon, Non-equilibrium Green's function (NEGF) has proposed as a strong method to investigate electrical properties \cite{pourfath2014non}. Tight-binding approach along with NEGF has been used to study carrier transport in the presence of disorders. 

Charged impurity in the substrate have been reported as the main source of scattering even for the cleanest Graphene devices \cite{hwang07,ando06}. Impurity scattering in Graphene on SiO$_2$ is not due to point defects presented in the parent material, but rather is likely caused by charged impurities in the SiO$_2$ substrate \cite{adam07,chen08charged}. For high quality SiO$_2$, an impurity concentration as high as $\mathrm{10^{11} cm^{-2}}$ is always reported. Tan et al. [\cite{tan07}] have examined many Graphene devices and have shown that the large differences in transport properties due to the charged impurities existed on the SiO$_2$ substrates. Specifically, they have found the impurity concentration of studied devices varies between $\mathrm{2-15 \times 10^{11} cm^{-2}}$. Adam et al. [\cite{adam07}] divided substrates in two groups, very clean substrate (substrate with impurity concentration in the range of $\mathrm{10 \times 10^{10}cm^{-2}}$) and very dirty substrate (substrate with impurity concentration in the range of $\mathrm{350 \times 10^{10} cm^{-2}}$). BN substrate is experimentally compared with SiO$_2$ and it is observed that charged impurity density in BN is approximately one order of magnitude lower than it in SiO$_2$ \cite{burson13}.

Multi-orbital Hamiltonian (s, p$_x$, p$_y$, p$_z$) is considered for modeling of the devices. Device Hamiltonian can be written as:
\begin{equation}
H=\sum_{i,\alpha} \epsilon_{\alpha}c_{i,\alpha}^\dagger c_{i,\alpha}+\sum_{\left<i,j\right>;\alpha,\beta}t_{i,j;\alpha,\beta}c_{i;\alpha}^\dagger c_{j;\beta},
\end{equation}
that i and j stand for the atomic sites, $\alpha$ and $\beta$ are atomic orbitals. $\epsilon_{\alpha}$ is on-site potential corresponding to the s- and p-orbitals and $t_{i,j;\alpha,\beta}$ is the hopping matrix elements between $\alpha$ orbital at i and $\beta$ orbital at j atomic sites. The $c_{i;\alpha}^\dagger$($c_{j;\beta}$) is creation (annihilation) operator. The TB parameters are taken from Ref. [\cite{tomanek88}]: $\epsilon_s$ = −7.3 eV, $\epsilon_p$ = 0.0 eV, $V_{ss\sigma}$ = −4.30 eV, $V_{sp\sigma}$ = 4.98 eV, $V_{pp\sigma}$ = 6.38 eV, $V_{pp\pi}$ = −2.66 eV.

It is assumed that charged impurities are randomly distributed with a density of $n_{imp}$. The potential has a Gaussian form that commonly used in the literatures \cite{lewenkopf08,bardarson07,klos09,babaee2019spin}:
\begin{equation}
U_{imp}(r_i)=\sum^{N_{imp}}_{j=1}U_j\exp\left(-\frac{|r_i - r_j|^2}{2\xi^2} \right),
\end{equation}
where $N_{imp}=N\times n_{imp}$ ($N$ is the number of atoms) is the number of the charged impurities. $\xi$ is the range of the potential. Strength amplitude $U_j$ taken from a uniform distribution over the interval $[-\delta U,\delta U]$ that is reported to be $\delta U\sim \mathrm{0.3 eV}$ \cite{hwang07,ando06}. $r_i$ and $r_j$ run over atom and impurity positions, respectively.

Surface roughness modulates positions and directions of atomic orbitals. Hopping parameters are modulated with using Harrison's model $t_{ij}\propto 1/d^2~$ \cite{harrison2012electronic} along with Slater-Koster approach \cite{slater54} where $d$ is bonding length after applying surface roughness. Substrate's surface roughness is a statistical phenomenon which can be described with a Gaussian autocorrelation function (ACF)~\cite{goodnick85,ishigami07}:
\begin{equation}	
R(x,y)= \mathit{\delta h}^2\exp\left(-\frac{\mathit{x}^2}{\mathit{L_x}^2}-
\frac{\mathit{y}^2}{\mathit{L_y}^2} \right)\ .
\label{eq:corrugation}
\end{equation}
where $\mathit{L_x}$ and $\mathit{L_y}$ are the roughness correlation lengths in the $x$ and $y$-direction, respectively. $\mathit{\delta h}$ is the root mean square of the height fluctuation. To generate surface corrugation in the spatial domain, the auto-correlation function is Fourier transformed to obtain the spectral function. A random phase with even parity is applied and followed by an inverse Fourier transformation \cite{touski13}.

NEGF is used to study the electron transport in Graphene. The Green's function of the channel at each energy (E) is given by:
\begin{equation}
G(E)=\left[E-H-U_{imp}-\Sigma_S-\Sigma_D-\Sigma_{ph}\right],
\end{equation}
where $H$ is the Hamiltonian of the channel after modulation of surface roughness. $\Sigma_{S,D}$ states self-energy of the left- and right- contacts that can be expressed as:
\begin{equation}
\Sigma_{S,D} = \tau_{S,D}^{\dagger} g_{S,D}(\epsilon) \tau_{S,D}.
\end{equation} 
$g_{S,D}$ is the surface Green's function of the contacts that is obtained with highly converge Sancho's method \cite{sancho85}. $\Sigma_{ph}$ is self-energy of electron-phonon (e-ph) interaction that can be calculated by using lesser and greater Green's functions. The lesser and greater Green's functions for e-ph scattering can be written as \cite{koswatta07}:
\begin{align}
\begin{split}
\Sigma^{<,>}_{ph}(j,j;E) &= D_0 (n_w+1)G^{<,>}(j,j;E\mp hw_{op})\\
&+ D_0 n_w G^{<,>}(j,j;E\pm hw_{op}),
\end{split} 
\end{align}
where $D_0$ is the electron-phonon coupling constant and is calculated from electron-phonon interaction Hamiltonian. For carbon nanotube, $D_0$ is equal to 0.07 \cite{mahan03}. This value is used in many works for Graphene nanoribbon\cite{yoon11,akhavan12}. $n_w$ is phonon occupation number in thermal equilibrium. $hw_{op}$ is the energy of surface optical phonon that its amount for the different substrate is listed in Table. \ref{table:SOP}. The electron and hole correlation functions, $G^<$ and $G^>$, are given by:
\begin{equation}
G^{<,>}=G\Sigma^{<,>} G^\dagger,
\end{equation}
where 
\begin{equation}
\Sigma^{<,>} = \Sigma^{<,>}_S + \Sigma^{<,>}_D + \Sigma^{<,>}_{ph},
\end{equation}
The lesser and greater Green's functions for two leads are calculated by:
\begin{align}
\begin{split}
\Sigma^{<}_{S,D}&=i\Gamma_{S,D}(E)f_{S,D}(E)\\
\Sigma^{>}_{S,D}&=-i\Gamma_{S,D}(E)\left(1-f_{S,D}(E)\right)
\end{split} 
\end{align}
where $f_{S,D}(E)$ is Fermi-Dirac distribution function for source and drain contacts. The broadening for two leads can be obtained from self-energy with: 
\begin{equation}
\Gamma_{S,D}(E)=-i\left(\Sigma_{S,D} - \Sigma^{\dagger}_{S,D}\right).
\end{equation}
The imaginary part of the electron-phonon self-energy can be obtained from lesser and greater Green's functions as:
\begin{equation}
\Sigma^i_{ph}=-\frac{i}{2}\Gamma_{ph}(E)=\frac{1}{2}\left[\Sigma^{>}_{ph}-\Sigma^{<}_{ph}\right].
\end{equation}
The real part of self-energy is manifested as a shift of energy levels and is computed by using the Hilbert transform. Because of the small effect of the real part of electron-phonon self-energy, it can be neglected in order to simplify the computations \cite{koswatta07}. 

In the case of surface optical phonon, the above equations should be solved self-consistently to calculate the current. For incoherent situation, the current can be obtained by:
\begin{equation}
I = \frac{q}{h}\int_{-\infty}^{+\infty}\left(\mathrm{Trace}\left[
i\Gamma_S G^< \right] - \mathrm{Trace}\left[i\Sigma_S^{<}A\right]\right)dE,
\end{equation}
But for coherent transport (without electron-phonon scattering), the transmission can be obtained with the following equations.
\begin{equation}
T(E)=\mathrm{Trace}\left[G\Gamma_S G^\dagger \Gamma_D\right].
\end{equation}
Current and transmission for the incoherent and coherent situation are calculated, respectively. Charged impurity and surface roughness are two random mechanisms scattering. Because of this, many equivalent samples for each value of the disorder (surface roughness and charged impurity) is simulated, and then the average of results (transmission, conductivity, mobility and mean free path) is reported for each ensemble.

\section{Results and discussion}

\begin{figure}[t]
	\centering
	\includegraphics[width=0.9\columnwidth]{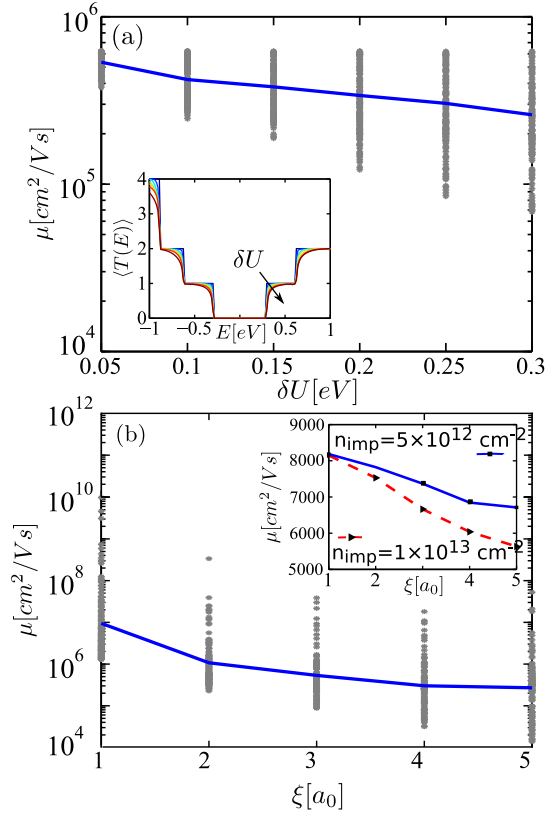}
	\caption{ Mobility as a function of (a) charged impurity strength and (b) range of potential. Inset of the Fig. (a) shows transmission as a function of energy for various $\delta U$. Mobility is plotted for two different charge impurity densities. Number of atoms in width (nW) is considered 15 for both of them.}
	\label{f:imp}
\end{figure}

\begin{figure*}[!t]
	\centering
	\includegraphics[width=2\columnwidth]{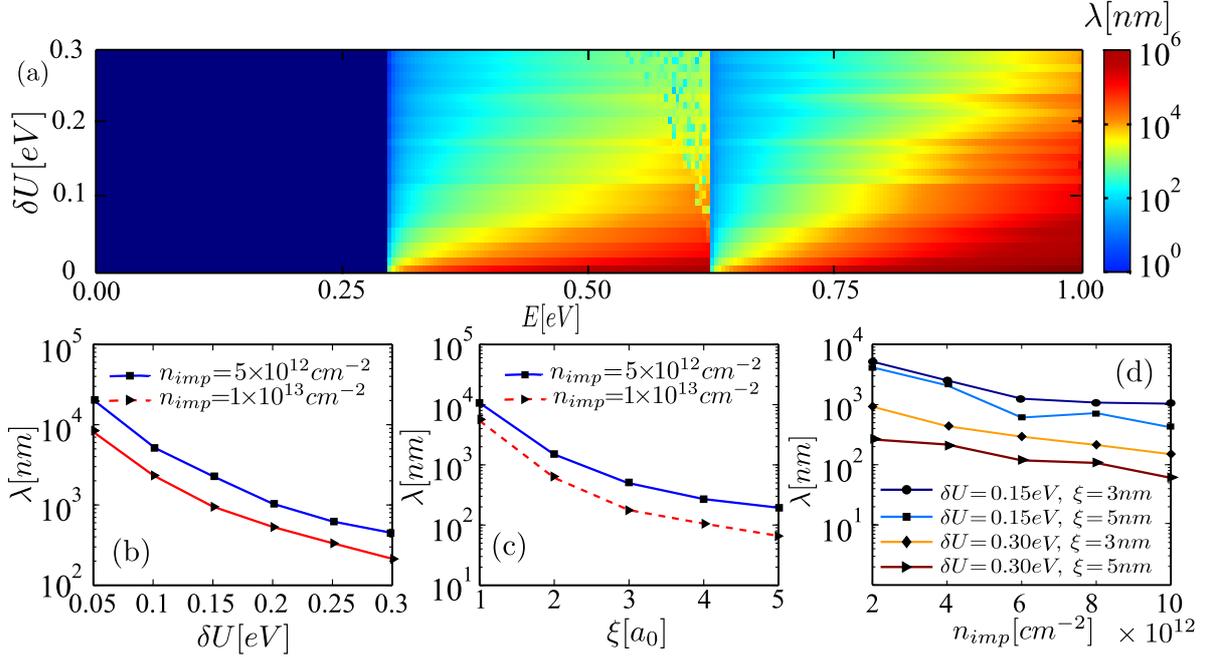}
	\caption{(a) Two-dimensional plot of the mean free path at varying energy and charged impurity strength. Mean free path is shown as a function of (b) $\delta U$ and (c) $\xi$ for two different charge impurity densities ($5\times 10^{12} cm^{-3}$ and $1\times 10^{13}cm^{-3}$).}
		. (d) Mean free path is shown as a function of charge impurity.
	\label{f:mfp}
\end{figure*}

\subsection{Charged impurity}
It is deeply accepted that the mobility-limiting factor in Graphene is the Coulomb scattering of charged impurities that reside either on Graphene or in the underlying substrate. We start by studying the electrical properties of supported Graphene in the presence of charged impurities. The mobility can be obtained with using $\mu=\sigma/ne$ that $\sigma$ is conductivity, $n$ is electron density and $e$ is unit of electrical charge. Obtained resistance with using NEGF can be partitioned into contact resistance ($R_B$) and channel resistance $R_{ch}$ \cite{ouyang11}. We subtracted interface resistivity from total resistivity, therefore, Mobility can be calculated using:
\begin{equation}
\mu=\frac{\sigma}{ne}=\frac{1}{(1/G-1/G_B)}\frac{L}{W}\frac{1}{ne}
\end{equation}  
that $G$ and $G_B$  are the conductivity in diffusive and ballistic regimes, respectively. $L$ and $W$ are length and width of the channel, respectively. Using the Landauer approach, conductivity can be found in linear response with using \cite{ryndyk09}:
\begin{equation}
G=G_0\int T(E) \left(-\frac{\partial f(E-E_f)}{\partial E} \right)dE,
\end{equation}
which $G_0=2e^2/(2\pi \hbar)$ and $f(E-E_f)$ is Fermi-Dirac distribution function. The mobility with respect to the strength and the range of CI potential is plotted in Fig. \ref{f:imp}(a) and (b), respectively. Charged impurity-limited mobility shows small dependency on $\delta U$, and indicates high mobility in the presence of CI (see Fig. \ref{f:imp}(a)). Smooth transmission in the inset of Fig. \ref{f:imp}(a) also proves the weak scattering due to long-range charged impurity scattering. As can be seen in Fig. \ref{f:imp}(b), the mobility shows more dependency on the range of potential, so that, the mobility decreases more than one order of magnitude with increasing $\xi$ from $a_0$ to $3a_0$. It should be noted that $a_0$ is the atom distance for nearest neighbor atoms. The mobility for two different impurity densities is indicated in the inset of Fig. \ref{f:imp}(b). As one can see, the effect of impurity density on mobility increases with increasing $\xi$. This increasing effect arises from the localized states that are created at a high $\xi$ in the presence of high CI density. The localized states in the presence of CI has been reported in Ref. [\cite{lherbier08}]. Mobility declines with increasing $\xi$, however, the mobility remains higher than $10^5 cm^2/Vs$ with these impurity densities for all the range of the potentials. It can be stated that mobility does not highly dependent on the CI that is in agreement with the results of Ref. [\cite{evaldsson08}].

In the following, we calculated the mean free path (MFP) in the presence of CI to investigate its effects on the electrical properties. MFP can be obtained with: $T(E)=M(E)/\left(1+L/\lambda\right)$ that $M$ is the number of subbands in the particular energy \cite{yazdanpanah12}. MFP obtained as a function of impurity amplitude and results are plotted in Fig. \ref{f:mfp}. MFP increases with increasing energy, however, mean free path falls down at the edge of subbands because of high scattering due to the high density of state. Due to longer MFP in comparison with channel length in the presence of CI, electron approximately transports ballistically in the device with $L$=14nm and $L$=22nm. MFP as a function of $\xi$ has been shown in Fig. \ref{f:mfp}(c). We consider two high CI densities to indicate that CI has a weak effect on the MFP. In this figure, $\lambda$ shows a $1/n_{imp}$ dependency that is in agreement with diffusive transport. However, one can observe distortion at $\xi=5 a_0$ related to the creation of localized states for high $\xi$.
%
\subsection{Substrate surface roughness}
\begin{figure}[t]
	\centering
	\includegraphics[width=1.0\columnwidth]{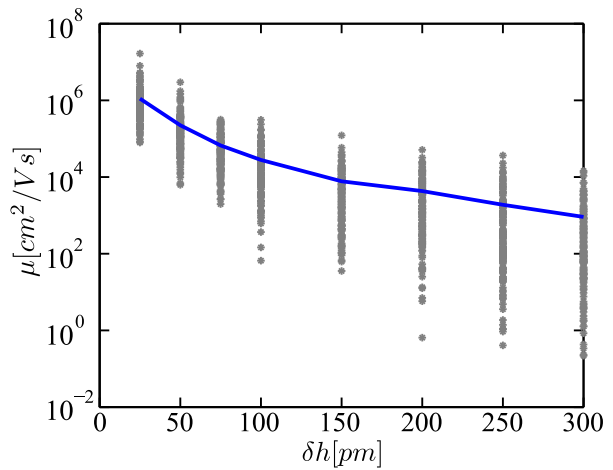}
	\caption{Plot of Mobility for each samples at the different surface roughness amplitude. Blue solid line indicate average of the mobility.}
	\label{f:sr}
\end{figure}

Substrate surface roughness is studied as another source of electron scattering. The surface roughness parameters for different substrates are summarized in our previous work \cite{chaghazardi16}. After the investigation of many samples, the mobility of each sample and its ensemble average is plotted as a function of surface roughness amplitude in Fig. \ref{f:sr}. Surface roughness-limited mobility ($\mu_{SR}$) decreases with increasing surface roughness but as it is obvious, the range of variation increases. Higher scattering due to surface roughness and the creation of localized states highly decline the mobility. So that, $\mu_{SR}$ decreases more than three orders of magnitudes when surface roughness amplitude increases from 30pm (BN's substrate surface roughness amplitude)\cite{xue11} to 250pm (SiO$_2$ substrate) \cite{ishigami07}. We observed the mobility remains more than $10^5 \mathrm{cm^2/Vs}$ for the investigated charged impurities, whereas, SiO$_2$ substrate's surface roughness scattering decreases mobility to $10^3 \mathrm{cm^2/Vs}$. This states that the effect of CI on mobility is negligible in comparison with the effect of surface roughness.

\begin{table*}
	\begin{center}
		\caption{Surface-optical phonon modes for various dielectric gates \cite{fischetti01,fratini08,konar10,perebeinos10}.
			\label{table:SOP}}
		\begin{tabular} {p{1.5cm}p{1.5cm}p{1.5cm}p{1.5cm}p{1.5cm}p{1.5cm}p{1.5cm}p{2.5cm}   }
			\hline
			\hline
			& HfO$_2$ & ZrO$_2$ & Al$_2$O$_3$ & SiO$_2$ & AlN    & BN   & SiC \\
			\hline
			$w_{SO,1}$ & 19.43   & 25.02   & 55.01       & 59.98   & 83.60  & 101.70 & 116 \\
			$w_{SO,2}$ & 52.87   & 70.80   & 94.29       & 146.51  & 104.96 & 195.70 & 167.58 \\
			\hline 
		\end{tabular}
	\end{center}
\end{table*}

\subsection{Surface optical phonon}
\begin{figure}[t]
	\centering
	\includegraphics[width=1.0\columnwidth]{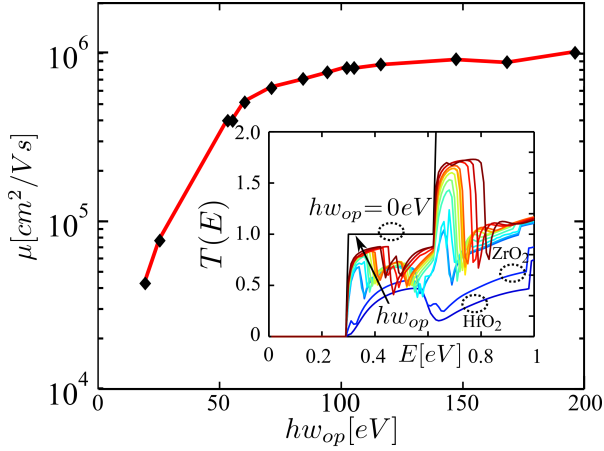}
	\caption{Mobility limited by surface optical phonon. Surface optical phonon is for various surface substrates that is reported in Table. \ref{table:SOP}. Inset figure shows transmission probability as a function of energy for different phonon energy. The arrow indicate the increasing of phonon energy ($\hbar w_{op}$).}
	\label{f:phonon}
\end{figure}

Surface optical phonon (SOP) is transferred to Graphene from underlying polar substrates. Phonon energy of the various substrates is reported in TABLE. \ref{table:SOP}. SOP-limited mobility ($\mu_{SOP}$) as a function of the reported surface optical phonon energy is shown in Fig. \ref{f:phonon}. Phonons with lower energy have a higher effect on mobility because high energy phonons only are effective at the high electric field. ZrO$_2$ and HfO$_2$ have the lowest optical phonon energies and the highest effects on $\mu_{SOP}$. As can be seen in this figure, these two substrates can decrease $\mu_{SOP}$ by one order of magnitude relative to other substrates. With other substrates (except ZrO$_2$ and HfO$_2$), mobility is approximately limited to $10^5 \mathrm{cm^2/Vs}$ that is in the range of $\mu_{CI}$.
We can't directly calculate the transmission probability in incoherent transport but we define effective transmission as $T(E)=I(E)/(f_S(E)-f_D(E))$. Inset figure of Fig. \ref{f:phonon} shows transmission probability versus energy. As mentioned before, the transmission of two substrates, ZrO$_2$ and HfO$_2$, is highly affected by surface optical phonon.

\subsection{Comparison of scattering disorders}
\begin{figure}[t]
	\centering
	\includegraphics[width=0.9\columnwidth]{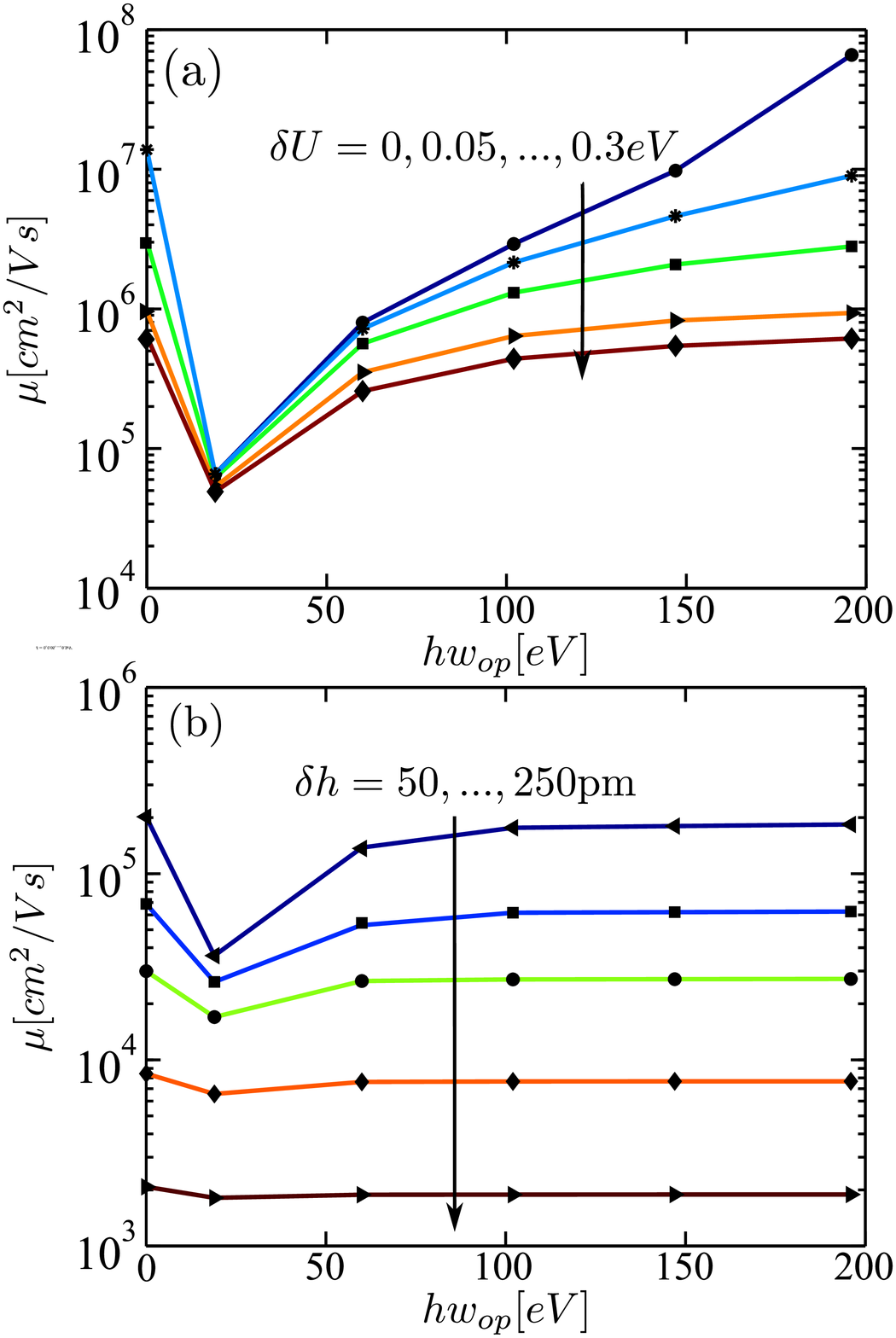}
	\caption{Mobility as a function of surface optical phonon energy for (a) different CI strength and (b) different surface roughness amplitudes.   Number of atoms in width (nW) is equal to 15 and $L_x=L_y=25nm$.}
	\label{f:phonon-imp-sr}
\end{figure}

In the following, the effects of surface optical phonon on the transport in the presence of charged impurity and surface roughness is studied. First, we analyze the effects of SOP in the presence of a charged impurity. The mobility versus the phonon energy is plotted for various $\delta U$ in Fig. \ref{f:phonon-imp-sr}(a). The mobility is limited by surface optical phonon at low phonon energy, whereas,  the effect of charged impurity increases with increasing phonon energy that can be explained by  Matthiessen's rule: $1/\mu = 1/\mu_{SOP}+1/\mu_{CI}$. We observed that the mobility is limited to $10^4 \mathrm{cm^2/Vs}$ with low phonon energies scattering, whereas, it is higher than $10^5 \mathrm{cm^2/Vs}$ with charged impurity scattering. As can be seen, the mobility doesn't depend on $\delta U$ for $hw_{OP}$=19.43 eV that is due to the high effect of SOP with this energy. 

We investigate the effects of both surface roughness and SOP. Fig. \ref{f:phonon-imp-sr}(b) shows the calculated mobility that has been limited by these mechanisms. Both scattering mechanisms contribute to limit the mobility at low surface roughness amplitudes. According to Matthiessen's rule, the mobility can be obtained with the contribution of two scattering mechanisms. However, as one can observe in Fig. \ref{f:phonon-imp-sr}(b), the mobility is independent on phonon energy for high surface roughness amplitudes.

\subsection{Comparing the results with experiment}
\begin{figure}[t]
	\centering
	\includegraphics[width=1.0\columnwidth]{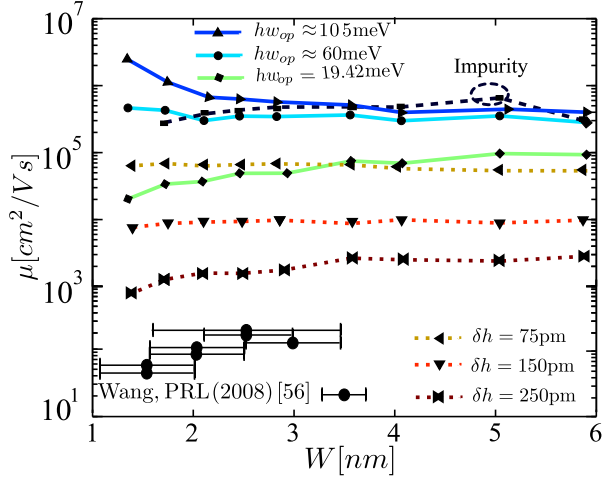}
	\caption{Mobility as a function of GNR width limited with various underlying substrate's disorders. For charged impurity we have $\delta U=0.3eV$, $\xi=3a_0$ and $n_{imp}=5\times 10^{12} cm^{-2}$. Three phonon energies ($hw_{op}=19.42eV$, $hw_{op}=59.98eV$ and $hw_{op}=101.7eV$) are considered. Also, three surface roughness amplitudes, $\delta h=75pm$, $\delta h=150pm$ and $\delta h=250pm$ are selected for BN and SiO$_2$ substrates. Solid lines, dashed line, and dotted lines indicate limited mobility by surface optical phonon, charged impurity and surface roughness, respectively.}
	\label{f:nw}
\end{figure}

Mobility as a function of nanoribbon width with considering three sources of scattering is plotted in Fig. \ref{f:nw}. $\delta U=\mathrm{300meV}$, $\xi=\mathrm{3a_0}$ and $n_{imp}=\mathrm{5 \times 10^12 cm^{-2}}$ are assumed for charged impurity. Charged impurity-limited mobility increases with increasing nanoribbon's width and decreases after $w=\mathrm{5nm}$ that it is in agreement with previous study \cite{betti11}. Surface optical phonon of three common substrates (HfO$_2 $, SiO$_2$ and BN) is considered. In addition, three surface roughness amplitudes are considered, $\delta h=\mathrm{75pm}$ for BN \cite{dean10}, $\delta h=\mathrm{150pm}$ and $\delta h=\mathrm{250pm}$ for SiO$_2$\cite{ishigami07,geringer09}. Mobility limited by impurity and SOP with $hw_{op}=$ 101.7 meV shows the highest values at different AGNR width. SiO$_2$ SOP ($hw_{op}=$59.98 meV) limits the mobility lower than BN's SOP but they approximately are in the same range, whereas, SOP-limited mobility for $hw_{op}=19.43 meV$ (HfO$_2$ dielectric) is one order of magnitude lower than them. This states that using of high-k dielectric (HfO$_2$ and ZrO$_2$) can decrease $\mu_{SOP}$ one order of magnitude for all of the width ranges, while the mobility highly limited by surface roughness. One can clearly observe that $\mu_{SR}$ is more than two orders of magnitudes lower than $\mu_{SOP}$ and $\mu_{CI}$. 

Mobility limited by various substrate's disorder is compared with experimental data from Wang et al \cite{wang08}. As one can observe, there is approximately one order of magnitude difference between our obtained mobility and reported mobility. Indeed, edge roughness is proposed as the dominant scattering on the limited mobility for narrow AGNR \cite{fang08,betti11}, but surface roughness is proposed dominant over edge roughness for wide AGNR\cite{touski13}.

Mobility limited by SOP due to BN substrate around $10^6 \mathrm{cm^2/Vs}$ that is one order of magnitude more than mobility limited by the surface roughness of BN substrate, whereas this difference with SiO$_2$ substrate ($\delta h\mathrm{=250pm}$ and $hw_{op}=$19.42 eV) is three orders of magnitudes. According to Matthiessen's rule, AGNR's mobility mainly limited by surface roughness for both substrates. Bischoff et al. [\cite{bischoff12}] proposed electrons transport through the localized charge puddles that may be created by surface roughness \cite{guinea08}. In opposite, V. Abramova et al. [\cite{Abramova13}] studied 50 Graphene nanoribbons on SiO$_2$ substrate and 12 GNRs on BN substrate. Mean average of GNR width on SiO$_2$ and BN substrates are reported $\mathrm{8.7nm}$ and $\mathrm{6.4nm}$, respectively. The mobility is reported $\mathrm{\mu =14cm^2/Vs}$ for GNR on SiO$_2$ substrate that is very smaller than our obtained results. However, a huge variation in mobility is observed for surface roughness, see Fig. \ref{f:sr}, so that the mobility is approximately distributed from $\mathrm{1}$ to $\mathrm{10^4 cm^2/Vs}$ for SiO$_2$ substrate ($\delta \mathrm{h=250pm}$). In addition, line edge roughness is the main source of scattering especially for narrow AGNR that is not considered here. This is proposed the substrate interaction as the main source of scattering for SiO$_2$ substrate. The minimum conductivity for GNR on SiO$_2$ and BN is compared. The minimum conductivity for SiO$_2$ is one order of magnitude smaller than GNR on BN in the room temperature and their difference can decline with temperature decreasing. N. JG. Couto [\cite{couto14}] theoretically and experimentally studied Graphene on the BN and they found random strain fluctuations are the dominant source of disorder that the random strain is created with surface roughness.

\section{Conclusion}
 The effects of the substrate disorders on the mobility of graphene nanoribbon was studied. We compared the effects of charged impurity, surface optical phonon and surface roughness. We find that mobility is mainly limited by surface roughness and somewhat by surface optical phonon. The effects of charged impurity on mobility can be neglected. The mobility in the presence of both charged impurity and surface optical phonon is investigated. We conclude that for low phonon energy, the mobility is limited by phonon, whereas, higher phonon energy has a smaller contribution to mobility. In the following, we studied the effects of both surface roughness and surface optical phonon. We observe that the mobility is mainly limited by surface roughness scattering. We find that the main sources of the scattering are surface roughness of substrate which can reduce mobility to around $\mathrm{1000 cm^2/Vs}$ for SiO$_2$ substrate.












\end{document}